\documentclass[preprint,showpacs]{revtex4}

\usepackage{graphicx}
\usepackage{amsmath}
\usepackage{amssymb}
\usepackage{dcolumn}
\usepackage{color}
\usepackage{bm}

\begin{document}


\title{Next-to-leading order QCD corrections to the top quark decay via
the Flavor-Changing Neutral-Current operators with mixing effects}

\author{Jia Jun Zhang}
\author{Chong Sheng Li}
 \email{csli@pku.edu.cn}
\author{Jun Gao}
\author{Hua Xing Zhu}
\affiliation{Department of Physics and State Key Laboratory of
Nuclear Physics and Technology, Peking University, Beijing
100871, China}
\author{C.-P. Yuan}
\email{yuan@pa.msu.edu} \affiliation{Department of Physics and
Astronomy, Michigan State University, East Lansing, Michigan,
48824-1116, USA}
\author{Tzu-Chiang Yuan}
\email{tcyuan@phys.sinica.edu.tw} \affiliation{Institute of Physics,
Academia Sinica, Nankang, Taipei 11529, Taiwan}

\date{\today}

\begin{abstract}
In this paper detailed calculations of the complete
$\mathcal{O}(\alpha_s)$ corrections to top quark decay widths
$\Gamma(t\rightarrow q+V)$ are presented ($V=g,\gamma,Z$). Besides
describing in detail the calculations in our previous paper
(arXiv:0810.3889), we also include the mixing effects of the
Flavor-Changing Neutral-Current (FCNC) operators for $t\rightarrow
q+\gamma$ and $t\rightarrow q+Z$, which were not considered in our
previous paper. The results for $t\rightarrow q+g$ are the same as
in our previous paper. But the mixing effects can either be large or small,
and increase or
decrease the branching ratios for $t\rightarrow q+\gamma$ and
$t\rightarrow q+Z$, depending on the values of the
anomalous couplings ($\kappa^{g,\gamma,Z}_{\mathrm{tq}}/\Lambda$,
$f^{g,\gamma,Z}_{\mathrm{tq}}$ and
$h^{g,\gamma,Z}_{\mathrm{tq}}$).
\end{abstract}

\maketitle

\section{Introduction}

Physics beyond the standard model (SM) can manifest itself by
altering the expected rates of Flavor-Changing Neutral-Current
(FCNC) interactions. Thus, testing SM and probing new physics
through top quark FCNC decay is interesting. The top quark FCNC
processes $t\rightarrow q+V$ ($V=g,\gamma,Z$) have tiny branching
ratios in the SM and are probably unmeasurable at the CERN Large
Hadron Collider (LHC) and future colliders. Therefore, any positive
signal of these rare decay events would imply new physics beyond the
SM. As the LHC will produce abundant top quark events (about $10^8$
per year), even in the initial low luminosity run
($\sim10\mathrm{fb}^{-1}/\mathrm{year}$) $8\times10^6$ top-quark
pairs and $3\times10^6$ single top quarks will be produced yearly.
Thus one may anticipate to discover the first hint of new physics by
observing anomalous couplings in the top-quark sector.

Recently, from their measurements of the total cross sections, both
D0 \cite{Abazov:2007ev} and CDF \cite{Aaltonen:2008qr}
collaborations at the Fermilab Tevatron have searched for
nonstandard-model single top quark production and set upper limits
on the anomalous FCNC couplings $\kappa^g_{\mathrm{tc}}/\Lambda$ and
$\kappa^g_{\mathrm{tu}}/\Lambda$, where the leading order (LO) cross
sections have been scaled to next-to-leading order (NLO)
\cite{Liu:2005dp} (or resummation \cite{Yang:2006gs}) predictions.
At the LHC, ATLAS collaboration has presented its sensitivity by
studying FCNC top decays \cite{Carvalho:2007yi}. These results show
that top quark FCNC couplings will provide a good probe to new
physics beyond the SM. Although there are many discussions in the
literatures on top quark production and rare decay processes
involving model-independent top quark FCNC couplings
\cite{Han:1996ep, Han:1996ce, Beneke:2000hk, Han:1995pk,
Hosch:1997gz, Obraztsov:1997if, Abe:1997fz, Han:1998tp, del
Aguila:1999ec, Chikovani:2000wi, Kidonakis:2003sc, Chekanov:2003yt,
AguilarSaavedra:2004wm}, most of them were based on LO calculations.
However, especially for $t\rightarrow q+g$, due to the large
uncertainties from the renormalization scale dependence in its LO
prediction through the strong coupling constant $\alpha_s$, it is
necessary to improve the theoretical prediction to NLO in order to
match the expected experimental accuracy at the LHC. Because of the
importance of NLO corrections for the experiments
\cite{Aaltonen:2008qr}, we calculated the NLO QCD corrections to the
partial decay widths and decay branching ratios of top quark FCNC
processes $t\rightarrow q+V\ (V=g, \gamma, Z)$ more than one year
ago \cite{Zhang:2008yn}.

In this paper, we describe in detail the calculations in
Ref.~\cite{Zhang:2008yn} and consider the FCNC operator mixing
effects which were ignored in Ref.~\cite{Zhang:2008yn}. This paper
is organized as follows. In section \ref{sect2}, we show the
relevant dimension-five operators and the corresponding LO results
in the top sector. Section \ref{sect3} is devoted to the detail of
the calculation presented in Ref.~\cite{Zhang:2008yn}. Section
\ref{sect4} deals with the evolution of anomalous couplings
$\kappa^g_{\mathrm{tq}}$. The analytic results for the mixing
effects can be found in section \ref{sect5}, while the numerical
results are presented in section \ref{sect6}.

\section{Leading Order Results\label{sect2}}

New physics effects involved in top quark FCNC processes can be
incorporated in a model-independent way, into an effective
Lagrangian which includes the dimension-five operators as listed
below \cite{Beneke:2000hk}
\begin{eqnarray}\label{lag}
\mathcal{L}^{\mathrm{eff}}&=&-\frac{e}{\sin2\theta_W}\sum_{q=u,c}
\frac{\kappa^Z_{\mathrm{tq}}}{\Lambda}\bar{q}\sigma^{\mu\nu}
(f^Z_{\mathrm{tq}}+ih^Z_{\mathrm{tq}}
\gamma_5)tZ_{\mu\nu}
-e\sum_{q=u,c}\frac{\kappa^{\gamma}_{\mathrm{tq}}}{\Lambda}\bar{q}\sigma^{\mu\nu}
(f^{\gamma}_{\mathrm{tq}}+ih^{\gamma}_{\mathrm{tq}}\gamma_5)tA_{\mu\nu}\notag\\
&&-g_s\sum_{q=u,c}\frac{\kappa^g_{\mathrm{tq}}}{\Lambda}\bar{q}\sigma^{\mu\nu}T^a
(f^g_{\mathrm{tq}}+ih^g_{\mathrm{tq}}\gamma_5)tG_{\mu\nu}^a+\mathrm{H.c}.
\end{eqnarray}
where $\Lambda$ is the new physics scale, $\theta_W$ is the
weak-mixing angle, and $T^a$ are the conventional Gell-Mann
matrices. The coefficients $\kappa^Z_{\mathrm{tq}},
\kappa^\gamma_{\mathrm{tq}}$ and $\kappa^g_{\mathrm{tq}}$ are
normalized to be real and positive, while $f^V_{\mathrm{tq}},
h^V_{\mathrm{tq}} (V=Z,\gamma,g)$ are complex numbers satisfying
$|f|^2+|h|^2=1$.

From the effective Lagrangian above, we obtain the following LO
partial decay width of the FCNC top decay in the $D=4-2\epsilon$
dimension,
\begin{eqnarray}
\Gamma_0^\epsilon(t\rightarrow
q+g)&=&\frac{8\alpha_sm_t^3}{3}
\left(\frac{\kappa^g_{\mathrm{tq}}}{\Lambda}\right)^2
C_{\epsilon},\notag\\
\Gamma_0^\epsilon(t\rightarrow q+\gamma)&=&2\alpha
m_t^3\left(\frac{\kappa^{\gamma}_{\mathrm{tq}}}{\Lambda}
\right)^2
C_{\epsilon},\notag\\
\Gamma_0^\epsilon(t\rightarrow q+Z)&=&\frac{\alpha
m_t^3\beta_Z^{4-4\epsilon}}{\sin^22\theta_W}
\left(\frac{\kappa^Z_{\mathrm{tq}}}{\Lambda}\right)^2
(3-\beta_Z^2-2\epsilon)
\frac{C_{\epsilon}}{1-\epsilon},
\end{eqnarray}
where the masses of light quarks $q$ ($q=u, c$) have been neglected,
and $\beta_Z=\sqrt{1-M_Z^2/m_t^2}$, $C_{\epsilon}=
\frac{\Gamma(2-\epsilon)}{\Gamma(2-2\epsilon)}
(\frac{4\pi\mu^2}{m_t^2})^{\epsilon}$. Here, we also define
$\Gamma_0(t\rightarrow q+V)=\Gamma_0^{\epsilon}(t\rightarrow q+V)|_{
\epsilon\rightarrow0}$, which are consistent with the results of
Refs.~\cite{Beneke:2000hk, Carvalho:2007yi}.

\section{Next-to-Leading Order Results \label{sect3}}

Below, we present our calculation in detail for the inclusive decay
width of the top quark, up to the NLO with the LO partonic process
denoted as $t\rightarrow q+g$. The results of $t\rightarrow
q+\gamma$ and $t\rightarrow q+Z$ are similar to the $t\rightarrow
q+g$, and will be presented all together.

\begin{figure}
\scalebox{0.7}{\includegraphics{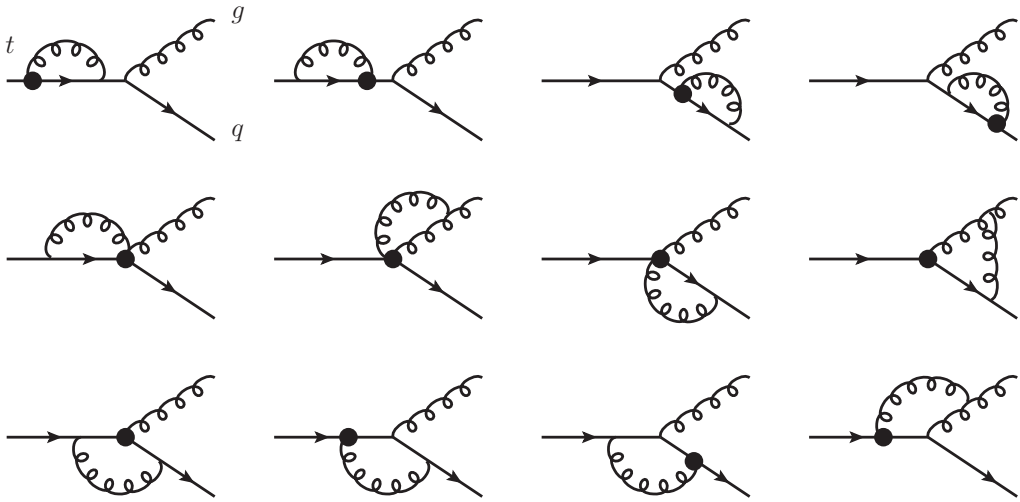}} \caption{\label{virt}
One-loop Feynman diagrams for $t\rightarrow q+g$. Here we use $q$ to
represents the up-quark and the charm-quark.}
\end{figure}

\begin{figure}
\scalebox{0.7}{\includegraphics{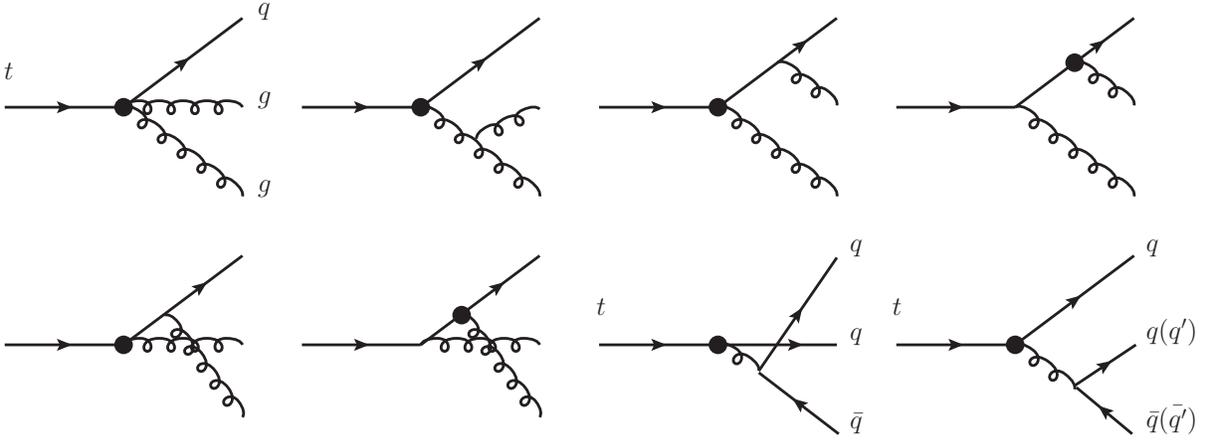}} \caption{\label{real} Feynman
diagrams of real gluon emission and gluon split. Here we use $q$ to
represents the up-quark and the charm-quark.}
\end{figure}

At the NLO, we need to include both one-loop virtual gluon
corrections (Fig.~\ref{virt}) and real gluon emission contribution
(Fig.~\ref{real}). We use dimensional regularization to regulate
both ultraviolet (UV) and infrared (IR) (soft and collinear)
divergences, with spatial-time dimension $D=4-2\epsilon$. The UV
divergences cancel after summing up the contributions from the
one-loop virtual diagrams and counterterms according to the same
convention used in Ref.~\cite{Liu:2005dp}. The soft divergences
cancel after adding up the virtual and real radiative corrections.
To cancel collinear singularities, we need to also include
contribution induced from gluon splitting to light quark pairs at
the same order in the QCD coupling.

The virtual correction of $t\rightarrow q+g$ contains UV and IR
divergences, which has the same form as in Ref.~\cite{Liu:2005dp}
and can be written as (the imaginary part is neglected):
\begin{eqnarray}
\mathcal{M}_{\mathrm{virt}}&=&\frac{\alpha_s}{12\pi}D_{\epsilon}
\left(-\frac{13}{\epsilon_{\mathrm{IR}}^2}-
\frac{17}{\epsilon_{\mathrm{IR}}}
+\frac{11}{\epsilon_{\mathrm{UV}}}+\frac{17}{3}\pi^2-15
\right)\mathcal{M}_0\notag\\
&&+\left(\frac{1}{2}\delta Z_2^{(g)}+\frac{1}{2}\delta
Z_2^{(q)}+\frac{1}{2}\delta Z_2^{(t)} +\delta Z_{g_s}+\delta
Z_{\kappa^g_{\mathrm{tq}}/\Lambda}\right)\mathcal{M}_0,
\end{eqnarray}
where
$D_{\epsilon}=\Gamma(1+\epsilon)[(4\pi\mu^2)/m_t^2]^{\epsilon}$, and
the UV divergences are renormalized by introducing counterterms for
the wave function of the external fields ($\delta Z_2^{(g)}, \delta
Z_2^{(q)}, \delta Z_2^{(t)}$) and the coupling constants ($\delta
Z_{g_s}, \delta Z_{\kappa^g_{\mathrm{tq}}/\Lambda}$). We define
these counterterms according to the same convention as in
Ref.\cite{Liu:2005dp}:
\begin{eqnarray}\label{RCs}
\delta
Z_2^{(g)}&=&-\frac{\alpha_s}{2\pi}D_{\epsilon}\left(
\frac{N_f}{3}-\frac{5}{2}\right)\left(
\frac{1}{\epsilon_{\mathrm{UV}}}
-\frac{1}{\epsilon_{\mathrm{IR}}}\right)-\frac{\alpha_s}{6\pi}D_{\epsilon}
\frac{1}{\epsilon_{\mathrm{UV}}},\notag\\
\delta Z_2^{(q)}&=&-\frac{\alpha_s}{3\pi}D_{\epsilon}\left(
\frac{1}{\epsilon_{\mathrm{UV}}}
-\frac{1}{\epsilon_{\mathrm{IR}}}\right),\notag\\
\delta
Z_2^{(t)}&=&-\frac{\alpha_s}{3\pi}D_{\epsilon}\left(
\frac{1}{\epsilon_{\mathrm{UV}}}+
\frac{2}{\epsilon_{\mathrm{IR}}}+4\right),\notag\\
\delta
Z_{g_s}&=&\frac{\alpha_s}{4\pi}\Gamma(1+\epsilon)(4\pi)^{\epsilon}
\left(\frac{N_f}{3}-\frac{11}{2}\right)
\frac{1}{\epsilon_{\mathrm{UV}}}+\frac{\alpha_s}{12\pi}
D_{\epsilon}\frac{1}{\epsilon_{\mathrm{UV}}},\notag\\
\delta
Z_{\kappa^g_{\mathrm{tq}}/\Lambda}&=&\frac{\alpha_s}{6\pi}\Gamma(1+\epsilon)
(4\pi)^{\epsilon}\frac{1}{\epsilon_{\mathrm{UV}}},\notag\\
\delta
Z_{\kappa^\gamma_{\mathrm{tq}}/\Lambda}&=&\delta
Z_{\kappa^Z_{\mathrm{tq}}/\Lambda}
=\frac{\alpha_s}{3\pi}
\Gamma(1+\epsilon)
(4\pi)^{\epsilon}\frac{1}{\epsilon_{\mathrm{UV}}},
\end{eqnarray}
where $N_f=5$ is the number of the active quark flavors. Here we
also present the necessary definition $\delta
Z_{\kappa^\gamma_{\mathrm{tq}}/\Lambda}$ and $\delta
Z_{\kappa^Z_{\mathrm{tq}}/\Lambda}$ used in the calculations of the
decay modes $t\rightarrow q+\gamma$ and $t\rightarrow q+Z$. Combine
these together and integrate through phase space, we get the virtual
contributions without UV divergences:
\begin{eqnarray}
\Gamma_{\mathrm{virt}}^g&=&\frac{\alpha_s}{6\pi}\Gamma_0^\epsilon
(t\rightarrow q+g)\left\{-\frac{13}{\epsilon_{\mathrm{IR}}^2}
+\frac{1}{\epsilon_{\mathrm{IR}}}\left[
-13\ln\frac{4\pi\mu^2}{m_t^2}+13\gamma_E+N_f-\frac{53}{2}\right]
\right.\notag\\
&&\left.+\left[-\frac{13}{2}\left(\ln\frac{4\pi\mu^2}{m_t^2}
-\gamma_E\right)^2-12\ln\frac{\mu^2}{m_t^2}
+\left(N_f-\frac{53}{2}\right)(\ln4\pi-\gamma_E)+\frac{55\pi^2}{12}-23
\right]\right\}.\notag\\
\end{eqnarray}
All the UV
divergences have canceled in $\Gamma^g_{\mathrm{virt}}$, as
they
must, but the infrared divergent piece is still present. We also
show the corresponding results for $t\rightarrow q+\gamma$ and
$t\rightarrow q+Z$ below,
\begin{eqnarray}
\Gamma_{\mathrm{virt}}^\gamma&=&\frac{\alpha_s}{3\pi}
\Gamma_0^\epsilon(t\rightarrow
q+\gamma)\left\{-\frac{2}{\epsilon_{\mathrm{IR}}^2}
+\frac{1}{\epsilon_{\mathrm{IR}}}\left[-2\ln\frac{4\pi\mu^2}
{m_t^2}+2\gamma_E-5\right]-2\ln\frac{\mu^2}{m_t^2}\right.\notag\\
&&\left.-\left(\ln\frac{4\pi\mu^2}{m_t^2}-\gamma_E\right)^2
+5\left(\gamma_E-\ln\frac{4\pi\mu^2}{m_t^2}\right)
-\frac{\pi^2}{6}-12\right\},\notag\\
\Gamma_{\mathrm{virt}}^Z&=&\frac{\alpha_s}{3\pi}
\Gamma_0^\epsilon(t\rightarrow
q+Z)\left\{-\frac{2}{\epsilon_{\mathrm{IR}}^2}
+\frac{1}{\epsilon_{\mathrm{IR}}}
\left[8\ln\beta_Z-2\ln\frac{4\pi\mu^2}{m_t^2}+2\gamma_E-5\right]-2\ln\frac{\mu^2}{m_t^2}\right.
\notag\\
&&+8\left(\ln\frac{4\pi\mu^2}{m_t^2}
-\gamma_E+\frac{3-2\beta_Z^2}{3-\beta_Z^2}\right)\ln\beta_Z
-\ln\frac{4\pi\mu^2}{m_t^2}\left(
\ln\frac{4\pi\mu^2}{m_t^2}-2\gamma_E+5\right)-8\ln^2\beta_Z\notag\\
&&\left.+4\mathrm{Li}_2\left(-\frac{1-\beta_Z^2}{\beta_Z^2}\right)
-\gamma_E^2+5\gamma_E-\frac{\pi^2}{6}-12\right\}.
\end{eqnarray}

The contribution from real gluon emission ($t\rightarrow q+g+g$) is
denoted as $\Gamma_{\mathrm{real}}(t\rightarrow q+g+g)$. In order to
cancel all the collinear divergences in the sum of virtual and real
radiative corrections, we also need to include the contributions
from gluon splitting into a pair of quark and antiquark in the
collinear region, which is denoted as
$\Gamma_{\mathrm{real}}(t\rightarrow q+q'+\bar{q}')$. Note that
there are two configurations of final state when the flavor of the
light quark coming from gluon splitting is the same as the light
quark directly from the FCNC coupling, and only one configuration
when they are different. The contributions of real gluon emission
($t\rightarrow q+g+g$) and gluon splitting ($t\rightarrow
q+q'+\bar{q}'$) are, respectively,
\begin{eqnarray}
\Gamma_{\mathrm{real}}(t\rightarrow q+g+g)&=
&\frac{\alpha_s}{6\pi}\Gamma_0^\epsilon(t\rightarrow q+g)\left[
\frac{13}{\epsilon_{\mathrm{IR}}^2}
+\frac{1}{\epsilon_{\mathrm{IR}}}\left(13\ln\frac{4\pi\mu^2}{m_t^2}
-13\gamma_E+\frac{53}{2}\right)\right.\notag\\
&&\left.+\frac{13}{2}\left(\ln\frac{4\pi\mu^2}{m_t^2}-\gamma_E\right)^2+
\frac{53}{2}\left(\ln\frac{4\pi\mu^2}{m_t^2}-\gamma_E\right)
-\frac{31}{4}\pi^2+\frac{171}{2}\right],\notag\\
\end{eqnarray}
and,
\begin{eqnarray}
\Gamma_{\mathrm{real}}(t\rightarrow q+q'+\bar{q}')&=
&\frac{\alpha_s}{6\pi}\Gamma_0^\epsilon(t\rightarrow q+g)\left[
-\frac{1}{\epsilon_{\mathrm{IR}}}N_f+N_f\left(\gamma_E
-\ln\frac{4\pi\mu^2}{m_t^2}-3\right)-\frac{1}{12}\right].\notag\\
\end{eqnarray}
After adding them together, the total real contributions can
be
written as
\begin{eqnarray}
\Gamma_{\mathrm{real}}^g&=&\Gamma_{\mathrm{real}}(t\rightarrow
q+g+g)+\Gamma_{\mathrm{real}}(t\rightarrow q+q'+\bar{q}')\notag\\
&=&\frac{\alpha_s}{6\pi}\Gamma_0^\epsilon(t\rightarrow
q+g)\left\{\frac{13}{\epsilon_{\mathrm{IR}}^2}
-\frac{1}{\epsilon_{\mathrm{IR}}}\left[-13\ln\frac{4\pi\mu^2}{m_t^2}+13\gamma_E
+N_f-\frac{53}{2}\right]\right.\notag\\
&&+\left[\frac{13}{2}\left(\ln\frac{4\pi\mu^2}{m_t^2}-\gamma_E\right)^2+\frac{53}{2}
\left(\ln\frac{4\pi\mu^2}{m_t^2}-\gamma_E\right)\right.\notag\\
&&\left.\left.-N_f\left(\ln\frac{4\pi\mu^2}{m_t^2}+3-\gamma_E\right)
-\frac{31\pi^2}{4}+\frac{1025}{12}\right]\right\}.
\end{eqnarray}
The corresponding real contributions of $t\rightarrow q+\gamma$ and
$t\rightarrow q+Z$ are also shown below
\begin{eqnarray}
\Gamma_{\mathrm{real}}^\gamma&=&\frac{\alpha_s}{3\pi}
\Gamma_0^\epsilon(t\rightarrow q+\gamma)\left\{
\frac{2}{\epsilon_{\mathrm{IR}}^2}+\frac{1}{\epsilon_{\mathrm{IR}}}
\left[2\ln\frac{4\pi\mu^2}{
m_t^2}-2\gamma_E+5\right]\right.\notag\\
&&\left.+\left(\ln\frac{4\pi\mu^2}{m_t^2}-\gamma_E\right)^2+5\left(
\ln\frac{4\pi\mu^2}{m_t^2}-\gamma_E\right)-\frac{7\pi^2}{6}+\frac{52}{3}\right\},\notag\\
\Gamma_{\mathrm{real}}^Z&=&\frac{\alpha_s}{3\pi}
\Gamma_0^\epsilon(t\rightarrow
q+Z)\left\{\frac{2}{\epsilon_{\mathrm{IR}}^2}
+\frac{1}{\epsilon_{\mathrm{IR}}}\left[-8\ln\beta_Z+2\ln\frac{4\pi\mu^2}{m_t^2}
-2\gamma_E+5\right]\right.\notag\\
&&+\left[4\left(4\ln\beta_Z+2\gamma_E-5\right)\ln\beta_Z
+\ln\frac{4\pi\mu^2}{m_t^2}
\left(\ln\frac{4\pi\mu^2}{m_t^2}-8\ln\beta_Z-2\gamma_E+5\right)+4\mathrm{Li}_2(\beta^2)\right.\notag\\
&&\left.\left.+\frac{4(1-\beta_Z^2)(1-4\beta_Z^2+\beta_Z^4)}{\beta_Z^4(3-\beta_Z^2)}
\ln(1-\beta_Z^2)
+\frac{12+135\beta_Z^2-43\beta_Z^4}{3\beta_Z^2(3-\beta_Z^2)}
+\gamma_E^2-5\gamma_E-\frac{11\pi^2}{6}\right]\right\}.\notag\\
\end{eqnarray}
Combine the real and virtual contributions, we obtain the full NLO
corrections for $t\rightarrow q+g$, $t\rightarrow q+\gamma$ and
$t\rightarrow q+Z$ as
\begin{eqnarray}\label{NLO}
\Gamma_{\mathrm{NLO}}(t\rightarrow
q+g)&=&\Gamma_{\mathrm{virt}}^g+\Gamma_{\mathrm{real}}^g\notag\\
&=&\frac{\alpha_s}{72\pi}\Gamma_0^\epsilon(t\rightarrow
q+g)\left[174\ln\left(\frac{\mu^2}{m_t^2}\right)
-12N_f\ln\left(\frac{\mu^2}{m_t^2}\right)-36N_f-38\pi^2+749\right],\notag\\
\Gamma_{\mathrm{NLO}}(t\rightarrow
q+\gamma)&=&\Gamma_{\mathrm{virt}}^{\gamma}+\Gamma_{\mathrm{real}}^{\gamma}\notag\\
&=&\frac{\alpha_s}{9\pi}\Gamma_0^\epsilon(t\rightarrow
q+\gamma)\left[-6\ln\left(\frac{\mu^2}{m_t^2}\right)-4\pi^2+16\right],\notag\\
\Gamma_{\mathrm{NLO}}(t\rightarrow
q+Z)&=&\Gamma_{\mathrm{virt}}^Z+\Gamma_{\mathrm{real}}^Z\notag\\
&=&\frac{\alpha_s}{3\pi}\Gamma_0^\epsilon(t\rightarrow
q+Z)\left[-2\ln\left(\frac{\mu^2}{m_t^2}\right)
+\frac{4(1-\beta_Z^2)(1-4\beta_Z^2+\beta_Z^4)}
{\beta_Z^4(3-\beta_Z^2)}\ln(1-\beta_Z^2)\right.\notag\\
&&+4\left(2\ln\beta_Z-\frac{9-\beta_Z^2}{3-\beta_Z^2}\right)\ln\beta_Z+4\mathrm{Li}_2(\beta_Z^2)
+4\mathrm{Li}_2\left(-\frac{1-\beta_Z^2}{\beta_Z^2}\right)\notag\\
&&\left.+\frac{12+135\beta_Z^2-43\beta_Z^4}{3\beta_Z^2(3-\beta_Z^2)}
-2\pi^2-12\right].
\end{eqnarray}
Note that the last expression in Eq.~\eqref{NLO} differs from the
Eq.~(9) in Ref.~\cite{Zhang:2008yn}, and the numerical difference is
about $2\%$, as shown below. 
But the conclusion on branching ratio for $t\rightarrow q+Z$ in
Ref.~\cite{Zhang:2008yn} is not changed, i.e., the NLO correction
is minuscule in branching ratio for $t\rightarrow q+Z$. 
Thus, up to the NLO, the partial decay width of the three FCNC
decays can be obtained
\begin{eqnarray}\label{TOTAL}
\Gamma(t\rightarrow q+V)&=&\Gamma_0(t\rightarrow
q+V)+\Gamma_{\rm NLO}(t\rightarrow q+V).
\end{eqnarray}
In order to study the effects of NLO corrections to the decay
branching ratios, we define the following branching ratios for later
numerical analysis:
\begin{eqnarray}
\mathrm{BR}_{\mathrm{LO}}(t\rightarrow
q+V)&=&\frac{\Gamma_0(t\rightarrow
q+V)}{\Gamma_0(t\rightarrow W+b)},\notag\\
\mathrm{BR}_{\mathrm{NLO}}(t\rightarrow
q+V)&=&\frac{\Gamma(t\rightarrow q+V)}{\Gamma(t\rightarrow W+b)}.
\end{eqnarray}
The decay width for the dominant top quark decay mode of
$t\rightarrow W+b$ at the tree level and the NLO can be found in
Ref.~\cite{Li:1990qf}, which we list below for the
convenience of the reader,
\begin{eqnarray}
\Gamma_0(t\rightarrow
W+b)&=&\frac{G_Fm_t^3}{8\sqrt{2}\pi}|V_{tb}|^2\beta_W^4(3-2\beta_W^2),
\notag\\
\Gamma(t\rightarrow W+b)&=&\Gamma_0(t\rightarrow W+b)\left\{1
+\frac{2\alpha_s}{3\pi}\left[2\left(\frac{(1-\beta_W^2)(2\beta_W^2-1)(\beta_W^2-2)}
{\beta_W^4(3-2\beta_W^2)}\right)\ln(1-\beta_W^2)\right.\right.\notag\\
&&\left.\left.-\frac{9-4\beta_W^2}{3-2\beta_W^2}\ln\beta_W^2
+2\mathrm{Li}_2(\beta_W^2)
-2\mathrm{Li}_2(1-\beta_W^2)-\frac{6\beta_W^4-3\beta_W^2-8}
{2\beta_W^2(3-2\beta_W^2)}-\pi^2\right]\right\},\notag\\
\end{eqnarray}
where $\beta_W\equiv\sqrt{1-m_W^2/m_t^2}$.

\section{Renormalization Group Equation Improvement\label{sect4}}

Due to the large scale dependence in the process $t\rightarrow q+g$,
we use the renormalization group evolution to improve the result of
perturbation theory. The anomalous couplings $\kappa^g$ satisfy
the following renormalization group equation
\begin{eqnarray}\label{RE}
\frac{\mathrm{d}\kappa^g}{\mathrm{d}\ln\mu}&=&-\gamma_{\kappa^g}
\kappa^g,
\end{eqnarray}
where
\begin{eqnarray}
\gamma_{\kappa^g}(g_s)&=&-2g_s^2\frac{\mathrm{d}Z_{\kappa^g,1}(g_s)}
{\mathrm{d}g_s^2}.
\end{eqnarray}
For simplicity, we neglect
the subscript tq of the anomalous couplings above and in the following
discussion. $Z_{\kappa^g,1}(g_s)$ is the residue of the renormalization
constant $\delta Z_{\kappa^g}$. Thus according to the Eq.~\eqref{RCs},
we have
\begin{eqnarray}
\gamma_{\kappa^g}(g_s)&=&-\frac{\alpha_s}{3\pi}.
\end{eqnarray}
Substitute it into Eq.~\eqref{RE}, we can solve the renormalization
group equation and get
\begin{eqnarray}
\kappa^g(\mu)&=&\kappa^g(\mu')\left(\frac{\alpha_s(\mu')}{
\alpha_s(\mu)}
\right)^{\frac{2}{3\beta_0}},
\end{eqnarray}
where $\beta_0=11-\frac{2}{3}N_f$. For $\alpha_s(\mu)$, we take it
by solving the following renormalization group equation
\begin{eqnarray}
\frac{\mathrm{d}\alpha_s(\mu)}{\mathrm{d}\ln\mu}&=&
2\beta(\alpha_s)=-\frac{\beta_0}{2\pi}\alpha_s^2.
\end{eqnarray}
We do not consider the higher order effects in the
$\gamma_{\kappa^g}$ and $\beta(\alpha_s)$ here because their
effects are small numerically.

\section{Contributions to $t\rightarrow q+\gamma$ and
$t\rightarrow q+Z$ from FCNC operator mixing \label{sect5}}

The NLO contributions given in Eq.~\eqref{NLO} are proportional to
the LO results. However, for the decay processes $t\rightarrow
q+\gamma$ and $t\rightarrow q+Z$, there are in general contributions
induced from the mixing of operators. Because we have no idea about
the magnitude of the coefficients $\kappa^V$, these operator mixing
contributions may be significant in some cases. In this section we
will present the contributions coming from the operator mixing so as
to complete the full $\mathcal{O}(\alpha_s)$ corrections to the
decay processes $t\rightarrow q+\gamma$ and $t\rightarrow q+Z$.

\begin{figure}
\scalebox{0.7}{\includegraphics{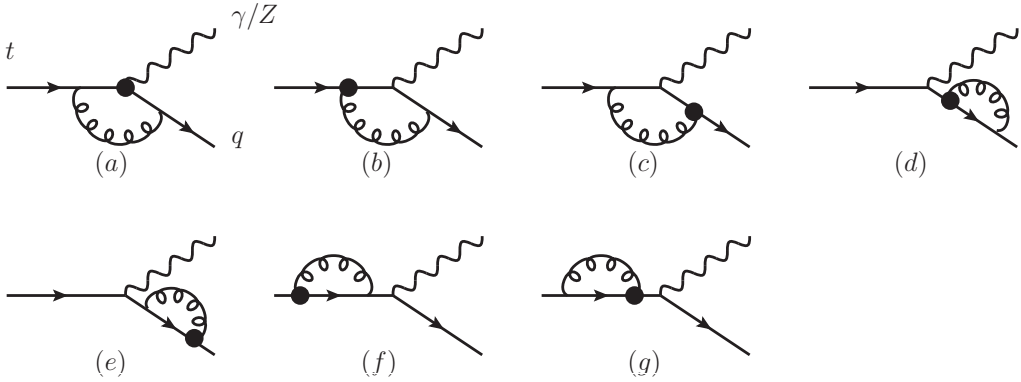}}
\caption{\label{virt_plus}Feynman diagrams of virtual corrections
for $t\rightarrow q+\gamma$ and $t\rightarrow q+Z$. Here we use
$q$ to represent the up-quark and charm-quark.}
\end{figure}

\begin{figure}
\scalebox{0.7}{\includegraphics{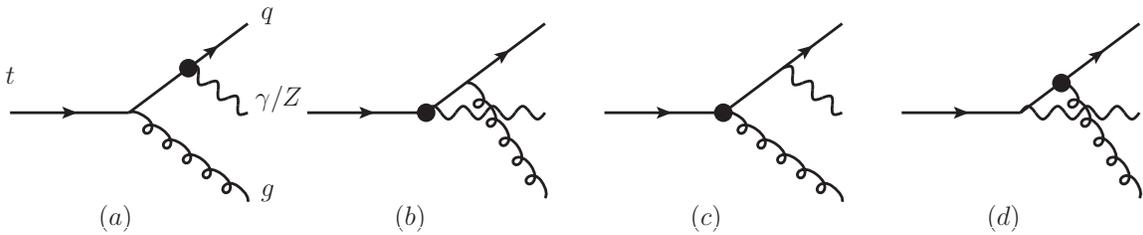}}
\caption{\label{real_plus}Feynman diagrams of real gluon emission
for $t\rightarrow q+\gamma$ and $t\rightarrow q+Z$. Here we use
$q$ to represent the up-quark and charm-quark.}
\end{figure}

The contributing Feynman diagrams are shown in Fig.~\ref{virt_plus}
and Fig.~\ref{real_plus}. In previous section, we consider only the
contributions from Fig.~\ref{virt_plus}(a) and
Fig.~\ref{real_plus}(a), (b). In case that all of the three
$\kappa^V$ are at the same order, the terms proportional to
$\kappa^g\kappa^\gamma$ and $\kappa^g\kappa^Z$ could contribute to
$t\rightarrow q+\gamma$ and $t\rightarrow q+Z$ with the same
significance as the ones we considered before. Thus, we will also
present these contributions below to investigate the effects from
operator mixing.

For convenience, we introduce the following abbreviations
\begin{eqnarray}
\Gamma_0^\gamma&=&2Q_f\alpha m_t^3
\left(\frac{\kappa_{\mathrm{tq}}^\gamma}{\Lambda}\right)\left(
\frac{\kappa_{\mathrm{tq}}^g}{\Lambda}\right)
C_\epsilon
(f^{\gamma*}_{\mathrm{tq}}f^g_{\mathrm{tq}}
+h^{\gamma*}_{\mathrm{tq}}h^g_{\mathrm{tq}}),\notag\\
\Gamma_0^Z&=&\frac{\alpha m_t^3\beta_Z^{-4\epsilon}}
{\sin2\theta_W\cos\theta_W}
\left(\frac{\kappa^Z_{\mathrm{tq}}}{\Lambda}\right)
\left(\frac{\kappa^g_{\mathrm{tq}}}{\Lambda}\right)
\frac{C_\epsilon}{1-\epsilon},\notag\\
S_3&=&\frac{s_3-Q_f\sin^2\theta_W}{\sin\theta_W},\notag\\
Q_F&=&Q_f\sin\theta_W,\notag\\
\omega_\beta &=& \sqrt{(1-\beta^2_Z)(3+\beta^2_Z)},\notag\\
c_1 &=& g_L^{Z*}g_L^gQ_F-g_R^{Z*}g_R^gS_3, \notag\\
c_2 &=& g_R^{Z*}g_R^gQ_F-g_L^{Z*}g_L^gS_3,
\end{eqnarray}
where $Q_f$ is the electric charge quantum number of the fermions,
e.g., $2/3$ for up-type quark and $-1/3$ for down-type quark. $s_3$
is the third component of the $\mathrm{SU}(2)_L$. Notice that we use
chiral parameters $g_L^i, g_R^i$ instead of $f^i_{\mathrm{tq}}$ and
$h^i_{\mathrm{tq}}$ in the $Z$-channel for simplicity, which have
the following relations,
$$
g_L^i=f^i_{\mathrm{tq}}-ih^i_{\mathrm{tq}},\qquad
g_R^i=f^i_{\mathrm{tq}}+ih^i_{\mathrm{tq}},\qquad
|g_L^i|^2+|g_R^i|^2=2.
$$

Denoting $\Gamma_{\mathrm{virt,mix}}^i (i=\gamma,Z)$ as
contributions to partial decay widths from the sum of
Fig.~\ref{virt_plus}(b)-(g) and the corresponding counter
terms, we obtain
\begin{eqnarray}
\Gamma_{\mathrm{virt,mix}}^\gamma&=&\Gamma_0^{\gamma}\left[\frac{2\alpha_s}{3\pi}
\left(-\frac{4}{\epsilon_{\mathrm{UV}}}+4\gamma_E-4
\ln\frac{4\pi\mu^2}{m_t^2}-11+\frac{2\pi^2}{3}\right)+ 2\delta
Z_{g\gamma}\right],\notag\\
\Gamma_{\mathrm{virt,mix}}^Z&=&\Gamma_0^Z\beta_Z^4(3-\beta_Z^2-2\epsilon)
\left\{
\frac{\alpha_s}{3\pi}c_1\left[-\frac{2}{\epsilon_{\mathrm{UV}}}
+2\gamma_E-2\ln\frac{4\pi\mu^2}{m_t^2}-4+\frac{2(1-\beta_Z^2)\ln(1-\beta_Z^2)-9}{3-\beta_Z^2}
\right.\right.\notag\\
&&\left.-\frac{4(2-\beta_Z^2)}{\beta_Z^2(3-\beta_Z^2)}
\sqrt{\frac{3+\beta_Z^2}{1-\beta_Z^2}}\tan^{-1}\left(
\frac{\omega_\beta}{1+\beta_Z^2}\right)
-\frac{8\overline{C_0}}{\beta_Z^4(3-\beta_Z^2)}\right]
+\frac{\alpha_s}{3\pi}c_2\left[-\frac{2}{\epsilon_{\mathrm{UV}}}\right.\notag\\
&&\left.\left.+2\gamma_E-2\ln\frac{4\pi\mu^2}{m_t^2}
+2\sqrt{\frac{3+\beta_Z^2}{1-\beta_Z^2}}
\tan^{-1}\left(\frac{\omega_\beta}{1+\beta_Z^2}\right)
-\frac{9-4\beta_Z^2}{3-\beta_Z^2}\right]+2\delta
Z_{gZ}\right\},
\end{eqnarray}
where $\overline{C_0}$ and the renormalization constants
$\delta
Z_{g\gamma}$ and $\delta Z_{gZ}$ are given below:
\begin{eqnarray}
\overline{C_0}&=&\ln\frac{
(1+\beta_Z^2)(2-2\beta_Z^2-\beta_Z^4)
+i\beta_Z^2(2+\beta_Z^2)\omega_\beta
}{2}\ln\frac{1+\beta_Z^2+i\omega_\beta}{2}\notag\\
&&
+2\mathrm{Li}_2(-\beta_Z^2)+\mathrm{Li}_2\left(\frac{
\beta_Z^2}{2}
\left(1+\beta_Z^2+i\omega_\beta
\right)\right)+
\mathrm{Li}_2\left(\frac{\beta_Z^2}{2}
\left(1+\beta_Z^2-i\omega_\beta
\right)\right)\notag\\
&&-\mathrm{Li}_2\left(\frac{\beta_Z^2}{2}
\left(-1+2\beta_Z^2+\beta_Z^4+i(1+\beta_Z^2)\omega_\beta
\right)
\right)\notag\\
&&-\mathrm{Li}_2\left(\frac{\beta_Z^2}{2}
\left(-1+2\beta_Z^2+\beta_Z^4-i(1+\beta_Z^2)\omega_\beta
\right)
\right),\notag\\
\delta
Z_{g\gamma}&=&\frac{\alpha_s}{3\pi}
\Gamma(1+\epsilon)(4\pi)^\epsilon
\frac{4}{\epsilon_{\mathrm{UV}}},\notag\\
\delta
Z_{gZ}&=&\frac{\alpha_s}{3\pi}
\Gamma(1+\epsilon)(4\pi)^\epsilon
\frac{c_1+c_2}{\epsilon_{\mathrm{UV}}}.
\end{eqnarray}
We can see that $\Gamma_{\mathrm{virt,mix}}^{\gamma}$
and
$\delta Z_{g\gamma}$ are consistent with Eqs. (4.8) and
(4.10) in
Ref.~\cite{Greub:1997hf}, up to a factor of $\frac{1}{3}$,
which is
absorbed into $Q_f$ in our formula. Besides,
$\Gamma_{\mathrm{virt,mix}}^Z$ and $\delta Z_{gZ}$ are
consistent
with $\Gamma_{\mathrm{virt,mix}}^{\gamma}$ and $\delta
Z_{g\gamma}$
if we take the limit $M_Z\rightarrow0$ and $Q_F=-S_3=1$.

It worths to briefly discuss the renormalization constant $\delta
Z_{gZ}$ here. To renormalize the operator involving the Z boson, we
need to introduce two renormalization constants: one for vector
current and another for axial vector current (or left-hand and
right-hand). Due to the similarity between the structure of the
anomalous couplings of $\gamma$ and $Z$ bosons in Eq.~\eqref{lag},
we expect that $\Gamma_{\mathrm{virt,mix}}^{\gamma}$ and
$\Gamma_{\mathrm{virt,mix}}^Z$ are equal in the limit of
$M_Z\rightarrow0$ and $Q_F=-S_3=1$. Thus, we introduce the
renormalization constants at the level of the decay width in
contrast to the usual practice which is done at the level of the
scattering amplitude. Nevertheless, this treatment would simplify
our calculation, so we take it as the definition of $\delta Z_{gZ}$
in this paper.

Denoting $\mathcal{M}^i_A (i=\gamma,Z)$ as the sum of
Fig.~\ref{real_plus} (a) and (b), and $\mathcal{M}^i_B (i=\gamma,Z)$
as the sum of Fig.~\ref{real_plus} (c) and (d), we can present the
contributions $\Gamma_{\mathrm{real,mix}}^i$ as
\begin{eqnarray}
\Gamma_{\mathrm{real,mix}}^\gamma&=&\frac{1}{m_t}\int
\sum\overline{|\mathcal{M}^{\gamma*}_A
\mathcal{M}^\gamma_B|}\mathrm{d}\Phi=\Gamma_0^\gamma\frac{
2\alpha_s}{9\pi }(2\pi^2-25),\notag\\
\Gamma_{\mathrm{real,mix}}^Z&=&\frac{1}{m_t}\int
\sum\overline{
|\mathcal{M}^{Z*}_A\mathcal{M}^Z_B|}
\mathrm{d}\Phi\notag\\
&=&\Gamma_0^Z\frac{\alpha_s}{3\pi}\left\{\frac{\beta_Z^2}{3}
\left[(-\beta_Z^4+21\beta_Z^2-72)c_1
+(24-21\beta_Z^2-\beta_Z^4)
c_2\right]\right.\notag\\
&&+2\omega_\beta[-(1+3\beta_Z^2)c_1
+(1-\beta_Z^2)c_2
]\left[\tan^{-1}\left(\frac{-1-\beta_Z^2}{\omega_\beta}\right)
+\tan^{-1}\left(\frac{1-\beta_Z^2}{\omega_\beta
}\right)\right]\notag\\
&&+(1-\beta_Z^2)\ln(1-\beta_Z^2)
[(9\beta_Z^2-21)c_1
+(5-\beta_Z^2)c_2]+8c_1
\notag\\
&&\left.\times\left[-\mathrm{Li}_2(\beta_Z^2)
+\mathrm{Li}_2\left(\frac{\beta_Z^2}{2}\left(1+\beta_Z^2+i
\omega_\beta\right)\right)
+\mathrm{Li}_2\left(\frac{\beta_Z^2}{2}
\left(1+\beta_Z^2-i
\omega_\beta\right)\right)\right]
\right\}
\end{eqnarray}

Combine these results together, we have
\begin{eqnarray}
\label{MIX2}
\Gamma_{\mathrm{mix}}^{\gamma}&=&\Gamma_{\mathrm{virt,mix}}^
{\gamma}
+\Gamma_{\mathrm{real,mix}}^{\gamma}=\Gamma_0^{\gamma}\frac{
4\alpha_s}{9\pi }\left(
6\ln\frac{m_t^2}{\mu^2}+2\pi^2-29\right),
\notag \\
\Gamma_{\mathrm{mix}}^Z&=&\Gamma_{\mathrm{virt,mix}}
^Z+\Gamma_{\mathrm{real,mix}}^Z\notag\\
&=&\frac{\alpha_s}{3\pi}\Gamma_0^Z\left\{\frac{\beta_Z^2}{3}
\left[c_1(-72-42\beta_Z^2+11\beta_Z^4)+c_2(24-48\beta_Z^2+11\beta_Z^4)
\right]\right.\notag\\
&&+(1-\beta_Z^2)[(2\beta_Z^4+9\beta_Z^2-21)c_1+(5-\beta_Z^2)c_2]
\ln(1-\beta_Z^2)\notag\\
&&+2\beta_Z^2\frac{\omega_\beta}{1-\beta_Z^2}
\tan^{-1}\left(\frac{\omega_\beta}{1+\beta_Z^2}\right)[
\beta_Z^2(3-\beta_Z^2)
c_2-2(2-\beta_Z^2)c_1]
\notag\\
&&+2\omega_\beta[-(1+3\beta_Z^2)
c_1+(1-\beta_Z^2)
c_2]\left[\tan^{-1}\left(\frac{-1-\beta_Z^2}{\omega_\beta}\right)
+\tan^{-1}\left(\frac{1-\beta_Z^2}{
\omega_\beta}\right)\right]
\notag\\
&&
+2\beta_Z^4(3-\beta_Z^2)(c_1+c_2)\ln\frac{m_t^2}{
\mu^2}
+8c_1X\bigg\},
\end{eqnarray}
where
\begin{eqnarray}
X&=&\mathrm{Li}_2\left(\frac{\beta_Z^2}{2}
\left(-1+2\beta_Z^2+\beta_Z^4+i(1+\beta_Z^2)
\omega_\beta\right)\right)+\mathrm{Li}_2\left(\frac{
\beta_Z^2}{2} \left(-1+2\beta_Z^2+\beta_Z^4-i(1+\beta_Z^2)
\omega_\beta\right)\right)\notag\\
&&-\mathrm{Li}_2(\beta_Z^2)-2\mathrm{Li}_2(-\beta_Z^2)
-\ln\frac{1+\beta_Z^2+i\omega_\beta}{2}\ln\frac{
(1+\beta_Z^2)
(2-2\beta_Z^2-\beta_Z^4)+i\beta_Z^2(2+\beta_Z^2)\omega_\beta
}{2}.\notag\\
\end{eqnarray}

\section{Numerical Analysis To Top FCNC Decay\label{sect6}}

\begin{table}
\caption{\label{table1}Numerical results of branching ratios. Here
$\mu=m_t$, $\kappa^V_{\mathrm{tq}}/\Lambda= 1\mathrm{TeV}^{-1}$,
$f^{\gamma*}_{\mathrm{tq}}f^g_{\mathrm{tq}}+h^{\gamma*}_{\mathrm{tq}}h^g_{\mathrm{tq}}=1$
and $g_R^{Z*}g_R^g=g_L^{Z*}g_L^g=1$.}
\begin{ruledtabular}
\begin{tabular}{cccccc}
BR&$\mathrm{BR_{LO}}$&$\mathrm{BR_{NLO}}$&$\mathrm{BR_{tot}}$
&$\mathrm{BR_{NLO}/BR_{LO}}$&$\mathrm{BR_{tot}/BR_{LO}}$
\\\hline
$t\rightarrow q+g$&1.0010&1.1964&-&1.195&-\\
$t\rightarrow q+\gamma$&0.0544&0.0542&0.0486&0.996&0.893\\
$t\rightarrow q+Z$&0.04484&0.04480&0.0459&0.999&1.025\\
\end{tabular}
\end{ruledtabular}
\end{table}
\begin{table}
\caption{\label{table2}Numerical results of the partial decay width.
Here $\mu=m_t$, $\kappa^V_{\mathrm{tq}}/\Lambda=
1\mathrm{TeV}^{-1}$,
$f^{\gamma*}_{\mathrm{tq}}f^g_{\mathrm{tq}}+h^{\gamma*}_{\mathrm{tq}}h^g_{\mathrm{tq}}=1$
and $g_R^{Z*}g_R^g=g_L^{Z*}g_L^g=1$.}
\begin{ruledtabular}
\begin{tabular}{cccccc}
Width[in unit
$\mathrm{GeV}$]&$\Gamma_{\mathrm{LO}}$&$\Gamma$&
$\Gamma_{\mathrm{tot}}$&$\Gamma/\Gamma_{\mathrm{LO}}$&
$\Gamma_{\mathrm{tot}}/\Gamma_{\mathrm{LO}}$\\\hline
$t\rightarrow q+g$&1.443&1.577&-&1.09&-\\
$t\rightarrow q+\gamma$&0.078&0.071&0.064&0.91&0.82\\
$t\rightarrow q+Z$&0.065&0.059&0.061&0.91&0.94\\
\end{tabular}
\end{ruledtabular}
\end{table}

For the numerical calculation of the branching ratios, we take the
SM parameters as given in Ref.~\cite{Amsler:2008zz}:
\begin{equation*}
m_t=171.2\mathrm{GeV},\qquad N_f=5,\qquad
m_W=80.398\mathrm{GeV},
\end{equation*}
\begin{equation*}
m_Z=91.1876\mathrm{GeV},\qquad\alpha=1/128,
\qquad\sin^2\theta_W=0.231,
\end{equation*}
\begin{equation*}
V_{\mathrm{tb}}=1, \qquad
G_F=1.16637\times10^{-5}\mathrm{GeV^{-2}}.\qquad
\end{equation*}
We analyze our results by choosing a special set of parameters and
fix $\mu=m_t$ in the following analysis unless specified.
Table~\ref{table1} and Table~\ref{table2} show the
$\mathcal{O}(\alpha_s)$ effects to various decay branching ratios
and partial decay widths, respectively, where
$\Gamma_{\mathrm{tot}}=\Gamma+\Gamma_{\mathrm{mix}}$ and
$\mathrm{BR}_{\mathrm{tot}}=
\Gamma_{\mathrm{tot}}/\Gamma(t\rightarrow W+b)$. From
Table~\ref{table1}, we see that the NLO correction increases the LO
branching ratio by about 20\% for $t\rightarrow q+g$, while the NLO
corrections are much smaller for the other two decay modes. But
after including the operator mixing effects, the branching ratio for
$t\rightarrow q+\gamma$ can decrease by about $10\%$, assuming
$\frac{\kappa^g_{\mathrm{tq}}}{\Lambda}=\frac{\kappa^{\gamma}_{\mathrm{tq}}}
{\Lambda}=1\mathrm{TeV}^{-1}$
and $f^{\gamma*}_{\mathrm{tq}}f^g_{\mathrm{tq}}
+h^{\gamma*}_{\mathrm{tq}}h^g_{\mathrm{tq}}=1$. For $t\rightarrow
q+Z$, the branching ratio can increase by about $2\%$, assuming
$\frac{\kappa^g_{\mathrm{tq}}}{\Lambda}=\frac{\kappa^Z_{\mathrm{tq}}}{\Lambda}
=1\mathrm{TeV}^{-1}$
and $g^{Z*}_Rg^g_R=g^{Z*}_Lg^g_L=1$. From Table~\ref{table2} we can
see that for the partial decay width the NLO results modify the LO
results by about $9\%$ in magnitude for all the three modes, and the
operator mixing effects can decrease the partial width by about
$9\%$ and increase by about $3\%$ with the above assumptions of
parameters for $t\rightarrow q+\gamma$ and $t\rightarrow q+Z$,
respectively.

\begin{figure}
\scalebox{1}{\includegraphics*[120,415][500,660]{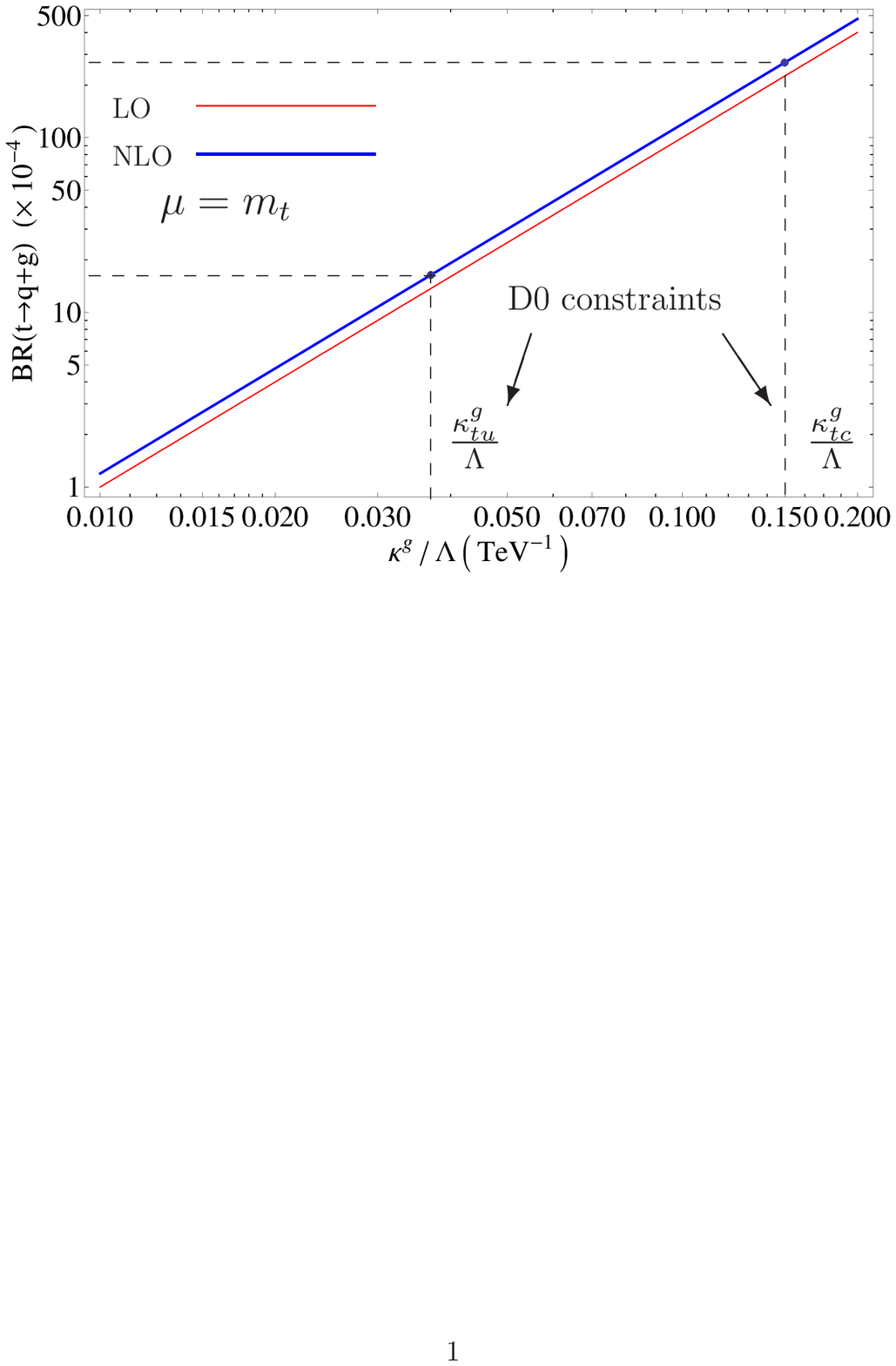}}
\caption{\label{BRvsKappa} Branching ratio as a function of
$\frac{\kappa^g}{\Lambda}$. Here $\mu=m_t$. We also give the D0
limits from Ref.~\cite{Abazov:2007ev}.}
\end{figure}

For convenience, we show the branching ratio of
$t\rightarrow q+g$
as a function of $\kappa^g/\Lambda$ in Fig.~\ref{BRvsKappa}. Using the upper limits measured
by the
D0 Collaboration at the Tevatron \cite{Abazov:2007ev}, we
get the
following results:
\begin{eqnarray}
\frac{\kappa^g_{\mathrm{tc}}}{\Lambda}<0.15\mathrm{TeV}^{-1}
&\Rightarrow&
\mathrm{BR}(t\rightarrow c+g)<2.69\times10^{-2},\notag\\
\frac{\kappa^g_{\mathrm{tu}}}{\Lambda}<0.037\mathrm{TeV}^{-1
}&\Rightarrow&
\mathrm{BR}(t\rightarrow u+g)<1.64\times10^{-3}.
\end{eqnarray}
Using our previous results \cite{Liu:2005dp, Yang:2006gs,
Zhang:2008yn}, CDF Collaboration presented a more precise
results
for the anomalous couplings and the branching ratios in a
recent
letter \cite{Aaltonen:2008qr}
\begin{eqnarray}
\frac{\kappa^g_{\mathrm{tc}}}{\Lambda}<0.069\mathrm{TeV}^{-1
}&\Rightarrow&
\mathrm{BR}(t\rightarrow c+g)<5.7\times10^{-3},\notag\\
\frac{\kappa^g_{\mathrm{tu}}}{\Lambda}<0.018\mathrm{TeV}^{-1
}&\Rightarrow&
\mathrm{BR}(t\rightarrow u+g)<3.9\times10^{-4}.
\end{eqnarray}

\begin{figure}
\scalebox{1}{\includegraphics*[120,400][485,650]{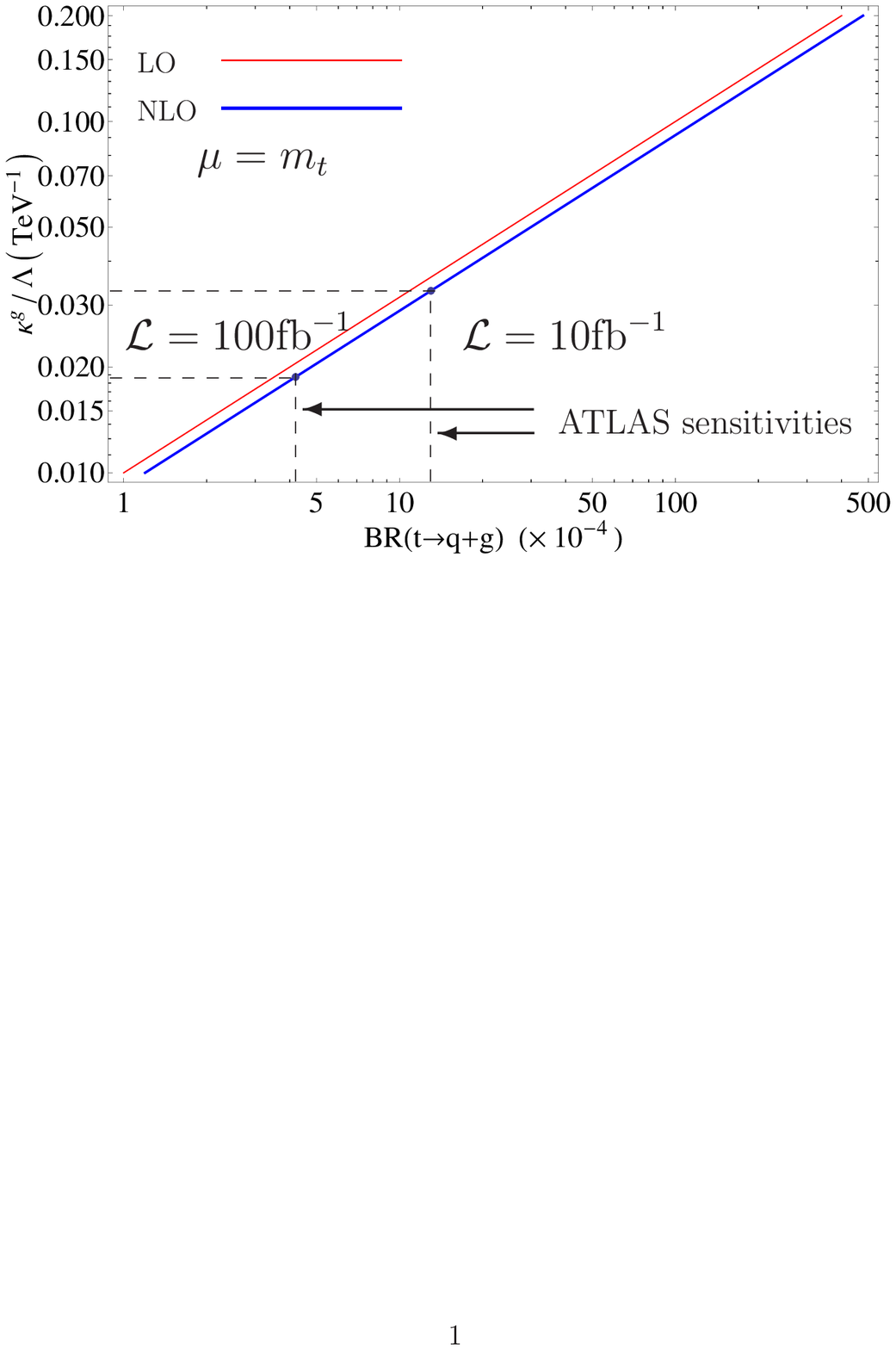}}
\caption{\label{KappavsBR} $\frac{\kappa^g_{tq}}{\Lambda}$ as a
function of branching ratio. Here $\mu=m_t$. We also show the ATLAS
sensitivities from Ref.~\cite{Carvalho:2007yi}.}
\end{figure}

Following the analysis in Ref.~\cite{Carvalho:2007yi}, we
plot the
coupling $\kappa^g_{\mathrm{tq}}/\Lambda$ as a function of
the
branching ratio in Fig.~\ref{KappavsBR}, where the ATLAS
sensitivities for the two different expected integrated
luminosities
are also exhibited. From Fig.~\ref{KappavsBR}, we can see
that
the NLO prediction improves the sensitivities of the LHC
experiments
to measuring the top quark FCNC couplings. With an
integrated
luminosity of $10\mathrm{fb}^{-1}$, the ATLAS experiment
sensitivities can be translated to the following relations
on FCNC
couplings:
\begin{eqnarray}\label{ATLASlimits1}
\mathrm{BR}(t\rightarrow q+g)>1.3\times10^{-3}&\Rightarrow&
\frac{\kappa^g_{\mathrm{tq}}}{\Lambda}>0.033\mathrm{TeV}^{-1
},\notag\\
\mathrm{BR}(t\rightarrow
q+\gamma)>4.1\times10^{-5}&\Rightarrow&
\frac{\kappa^\gamma_{\mathrm{tq}}}{\Lambda}>0.028\mathrm{TeV
}^{-1},\notag\\
\mathrm{BR}(t\rightarrow q+Z)>3.1\times10^{-4}&\Rightarrow&
\frac{\kappa^Z_{\mathrm{tq}}}{\Lambda}>0.083\mathrm{TeV}^{-1
},
\end{eqnarray}
and with an integrated luminosity of $100\mathrm{fb}^{-1}$,
they can
be translated to the more stringent relations:
\begin{eqnarray}\label{ATLASlimits2}
\mathrm{BR}(t\rightarrow q+g)>4.2\times10^{-4}&\Rightarrow&
\frac{\kappa^g_{\mathrm{tq}}}{\Lambda}>0.019\mathrm{TeV}^{-1
},\notag\\
\mathrm{BR}(t\rightarrow
q+\gamma)>1.2\times10^{-5}&\Rightarrow&
\frac{\kappa^\gamma_{tq}}{\Lambda}>0.015\mathrm{TeV}^{-1},
\notag\\
\mathrm{BR}(t\rightarrow q+Z)>6.1\times10^{-5}&\Rightarrow&
\frac{\kappa^Z_{\mathrm{tq}}}{\Lambda}>0.037\mathrm{TeV}^{-1
}.
\end{eqnarray}

\begin{figure}
\scalebox{1}{\includegraphics*[120,410][490,660]{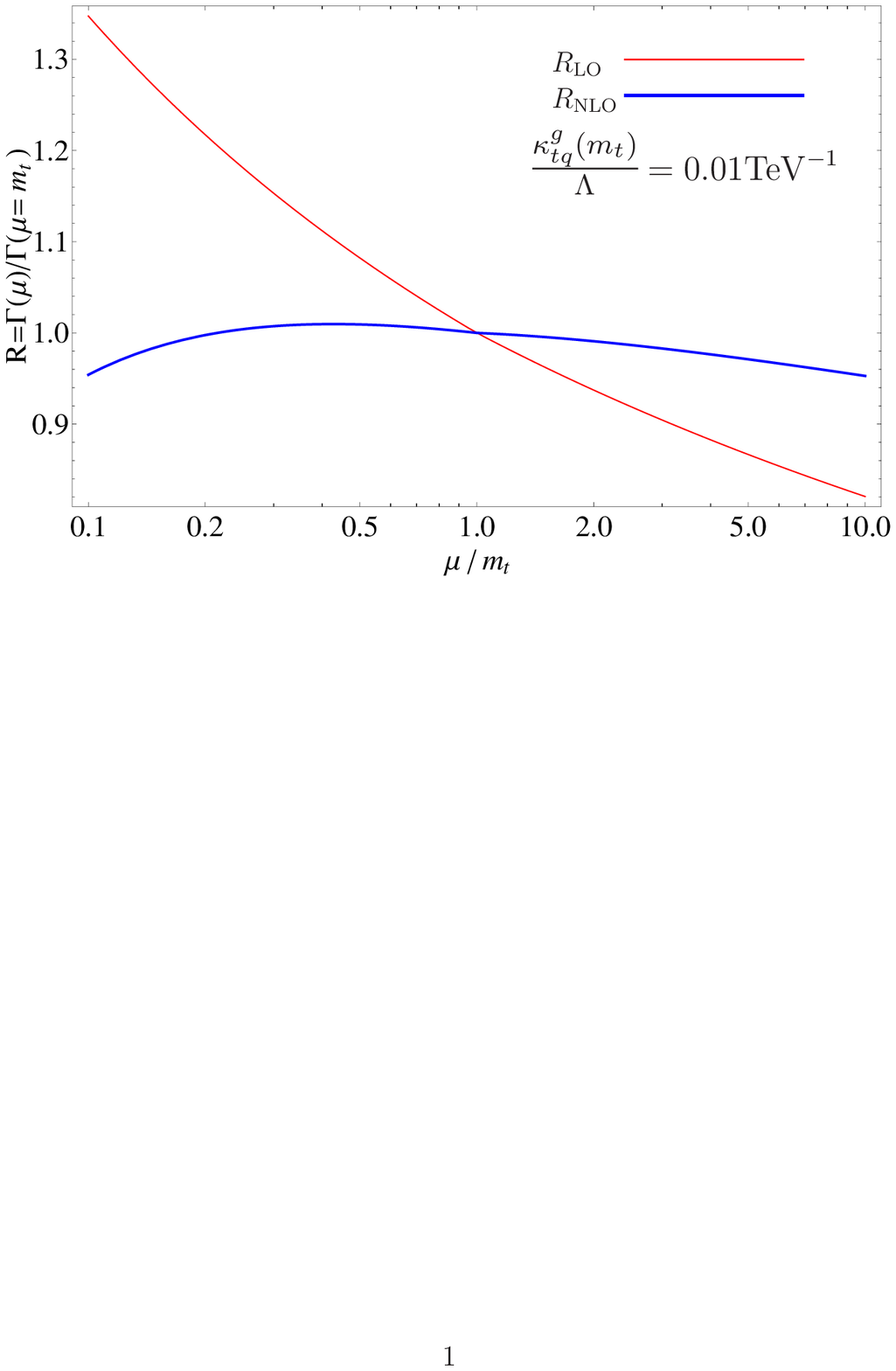}}
\caption{\label{RvsScale}The ratio R as a function of the
renormalization scale. Here,
$\frac{\kappa^g}{\Lambda}=0.01\mathrm{TeV}^{-1}$.}
\end{figure}

Finally, we illustrate the fact that the NLO prediction reduces the
theoretical uncertainty in its prediction on the decay branching
ratios and partial decay widths of the top quark. We define
$R_{\mathrm{LO}}(\mu)=\Gamma_0(\mu)/\Gamma_0(\mu=m_t)$ and
$R_{\mathrm{NLO}}(\mu)=\Gamma(\mu)/\Gamma(\mu=m_t)$, and show the
value of $R(\mu)$ as a function of $\mu$ for $t\rightarrow q+g$ in
Fig.~\ref{RvsScale}. It shows that the theoretical uncertainty from
the renormalization scale dependence can be largely reduced to a
couple of percent once the NLO calculation is taken into account, so
the NLO results give much more reliable theoretical predictions.

\begin{figure}
\begin{minipage}{0.45\textwidth}
\scalebox{0.5}{\includegraphics[150,210][550,650]{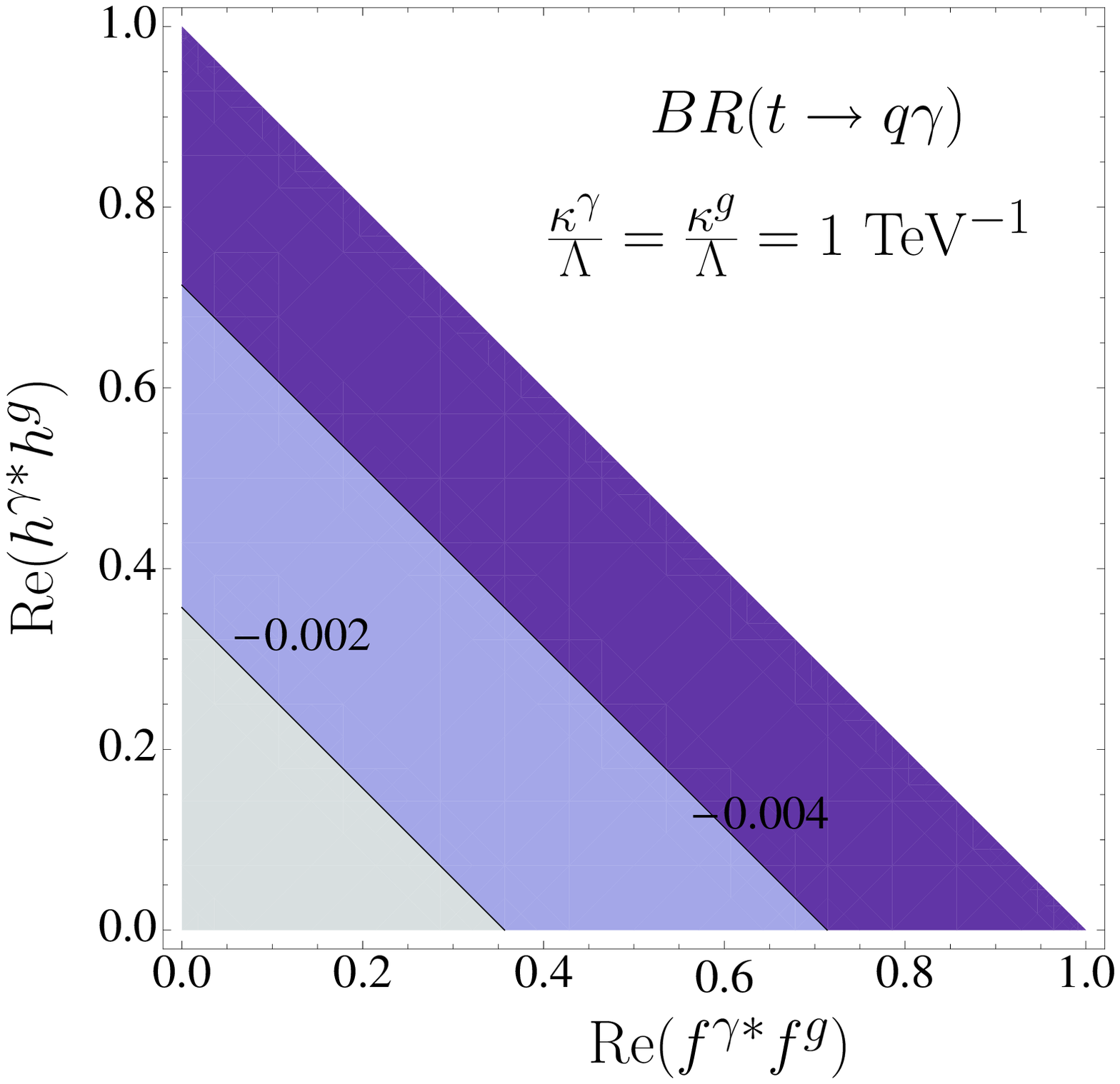}}
\end{minipage}\hfill
\begin{minipage}{0.45\textwidth}
\scalebox{0.5}{\includegraphics[150,210][550,650]{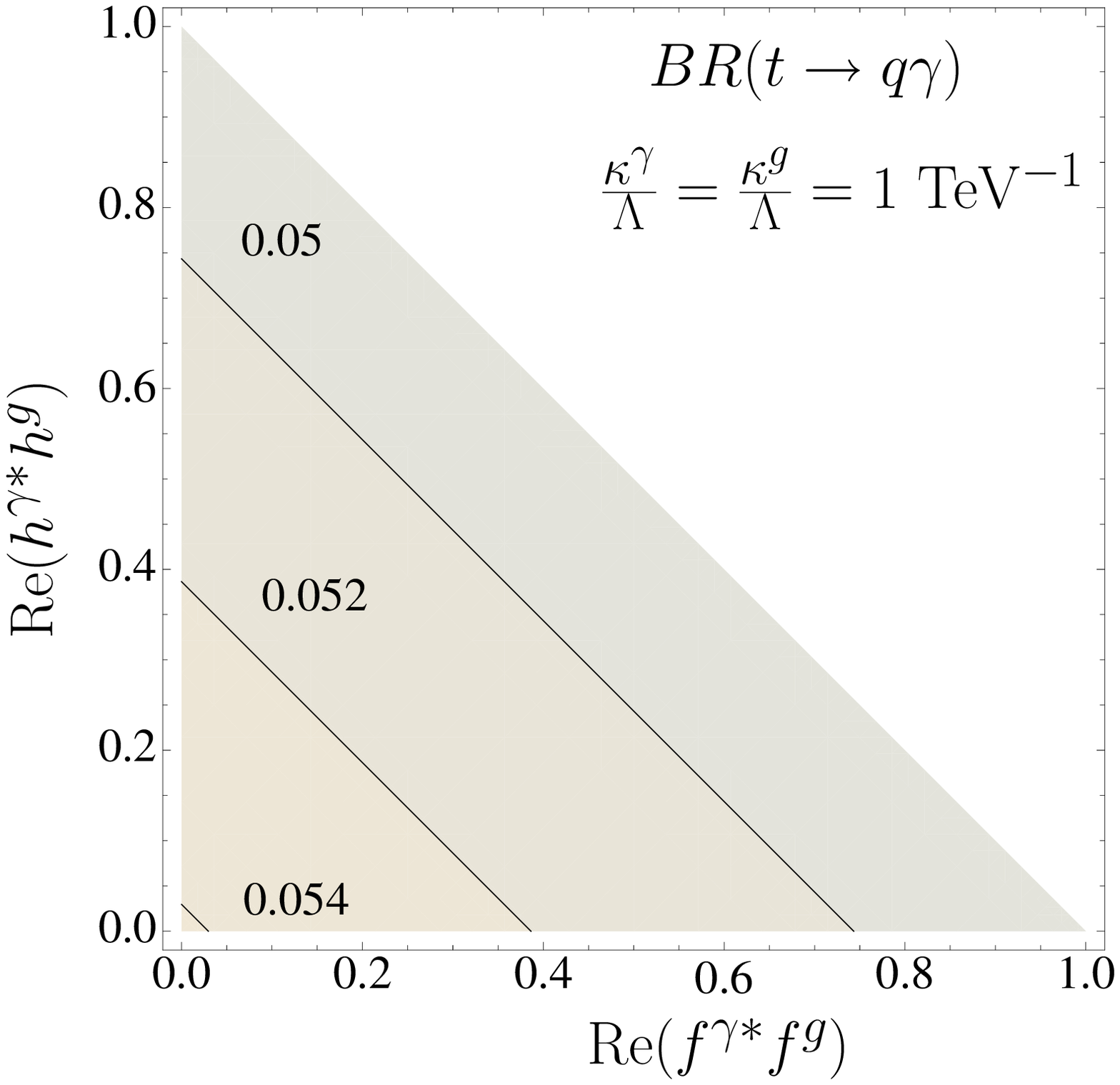}}
\end{minipage}\\
\begin{minipage}{0.45\textwidth}
\centering{(a)}
\end{minipage}\hfill
\begin{minipage}{0.45\textwidth}
\centering{(b)}
\end{minipage}
\caption{\label{gammamix} The contour curves of the decay branching
ratio $\rm BR(t\rightarrow q+\gamma)$ versus the variables
$\mathrm{Re}(f^{\gamma*}f^g)$ and $\mathrm{Re}(h^{\gamma*}h^g)$. We
show in (a) contributions induced from operator mixing effects; (b)
contributions including LO, NLO and mixing effects all together.
Here, we set
$\frac{\kappa^\gamma}{\Lambda}=\frac{\kappa^g}{\Lambda}=1\
\mathrm{TeV}^{-1}$ for simplicity.}
\end{figure}

\begin{figure}
\begin{minipage}{0.45\textwidth}
\scalebox{0.5}{\includegraphics[150,210][550,650]{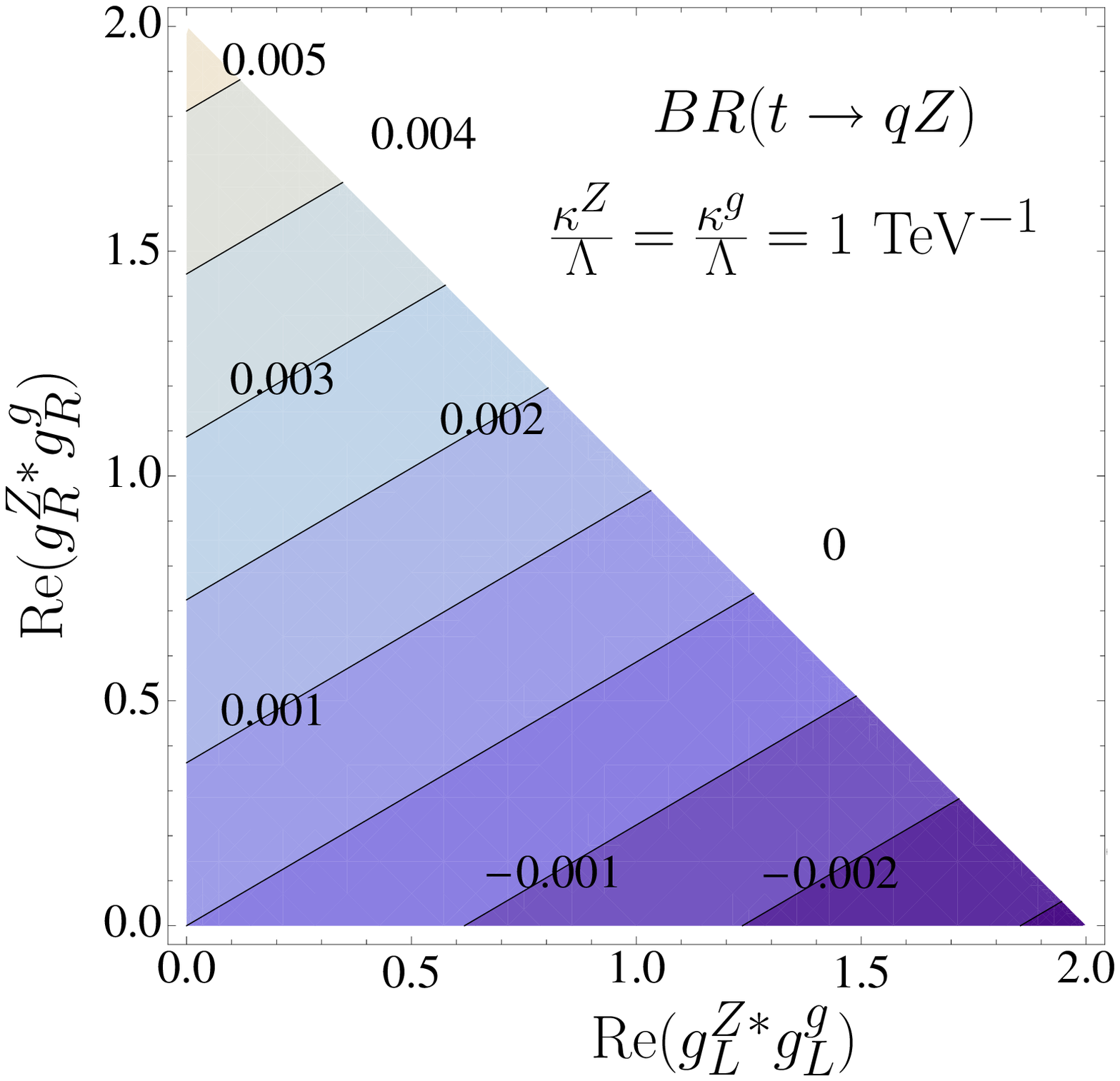}}
\end{minipage}\hfill
\begin{minipage}{0.45\textwidth}
\scalebox{0.5}{\includegraphics[150,210][550,650]{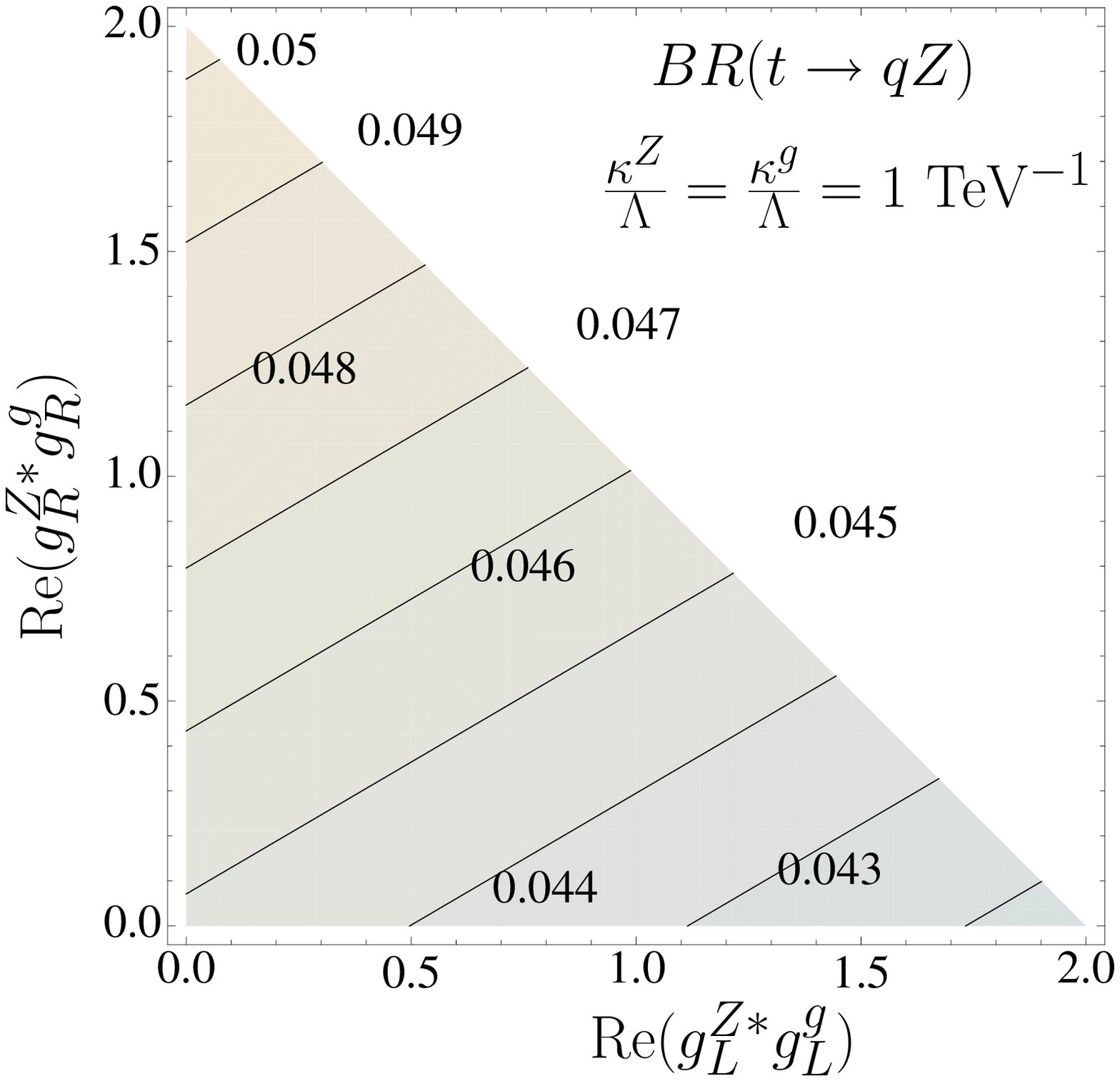}}
\end{minipage}\\
\begin{minipage}{0.45\textwidth}
\centering{(a)}
\end{minipage}\hfill
\begin{minipage}{0.45\textwidth}
\centering{(b)}
\end{minipage}
\caption{\label{Zmix} The contour curves of the decay branching
ratio $\rm BR(t\rightarrow q+Z)$ versus the variables
$\mathrm{Re}(g_L^{Z*}g_L^g)$ and $\mathrm{Re}(g_R^{Z*}g_R^g)$. We
show in (a) contributions induced from operator mixing effects; (b)
contributions including LO, NLO and mixing effects. Here, we set
$\frac{\kappa^Z}{\Lambda}=\frac{\kappa^g}{\Lambda}=1\
\mathrm{TeV}^{-1}$ for simplicity.}
\end{figure}

To investigate the contributions from the operator mixing effects,
we present the contour curves for the variables
$\mathrm{Re}(f^{\gamma*}f^g)$ (or $\mathrm{Re}(g_L^{Z*}g_L)$) and
$\mathrm{Re}(h^{\gamma*}h^g)$ (or $\mathrm{Re}(g_R^{Z*}g_R)$) for
the $\gamma$ (or $Z$) channel, where
$\frac{\kappa^\gamma}{\Lambda}=\frac{\kappa^g}{\Lambda}=1\
\mathrm{TeV}^{-1}$ (or $\frac{\kappa^Z}
{\Lambda}=\frac{\kappa^g}{\Lambda}=1\ \mathrm{TeV}^{-1}$). In
Fig.~\ref{gammamix} (a), we show the pure operator mixing effects to
the branching ratio of $t\rightarrow q+\gamma$, while in
Fig.~\ref{gammamix} (b) the total result are shown, which includes
the LO, NLO and mixing effects all together. Similarly, we present
the results for $t\rightarrow q+Z$ in Fig.~\ref{Zmix}.

Considering the mixing effects, we need to modify the inequalities
involving $\gamma$ and $Z$ in Eq.~\eqref{ATLASlimits1} and
Eq.~\eqref{ATLASlimits2}, which read as
\begin{eqnarray}
\lefteqn{\mathrm{BR}(t\rightarrow
q+\gamma)>4.1\times10^{-5}\Rightarrow}\notag\\
&&\left(\frac{\kappa^\gamma_{\mathrm{tq}}}{\Lambda}\right)^2
-0.1\left(\frac{\kappa^\gamma_{\mathrm{tq}}}
{\Lambda}\right)\left(\frac{\kappa^g_{\mathrm{tq}}}{\Lambda}
\right)
\mathrm{Re}(f^{\gamma*}_{\mathrm{tq}}f^g_{\mathrm{tq}}
+h^{\gamma*}_{\mathrm{tq}}h^g_{\mathrm{tq}})>7.6\times10^{-4
} \mathrm{TeV}^{-2},\notag\\
\lefteqn{\mathrm{BR}(t\rightarrow
q+Z)>3.1\times10^{-4}\Rightarrow}\notag\\
&&\left(\frac{\kappa^Z_{\mathrm{tq}}}{\Lambda}\right)^2+[
-3.6\mathrm{Re}(g_L^{Z*}g_L^g)
+6.2\mathrm{Re}(g_R^{Z*}g_R^g)]\left(\frac{\kappa^Z_{\mathrm
{tq}}}{\Lambda}\right)
\left(\frac{\kappa^g_{\mathrm{tq}}}{\Lambda}\right)\times10^
{-2}>6.9\times10^{-3} \mathrm{TeV}^{-2}\notag\\
\end{eqnarray}
for the integrated luminosity of $10\mathrm{fb}^{-1}$, and
\begin{eqnarray}
\lefteqn{\mathrm{BR}(t\rightarrow
q+\gamma)>1.2\times10^{-5}\Rightarrow}\notag\\
&&\left(\frac{\kappa^\gamma_{\mathrm{tq}}}{\Lambda}\right)^2
-0.1\left(\frac{\kappa^\gamma_{\mathrm{tq}}}{\Lambda}\right)
\left(\frac{\kappa^g_{\mathrm{tq}}}{\Lambda}\right)\mathrm{
Re}
(f^{\gamma*}_{\mathrm{tq}}f^g_{\mathrm{tq}}
+h^{\gamma*}_{\mathrm{tq}}h^g_{\mathrm{tq}})>2.2\times10^{-4
} \mathrm{TeV}^{-2},\notag\\
\lefteqn{\mathrm{BR}(t\rightarrow
q+Z)>6.1\times10^{-5}\Rightarrow}\notag\\
&&\left(\frac{\kappa^Z_{\mathrm{tq}}}{\Lambda}\right)^2+[
-3.6\mathrm{Re}(g_L^{Z*}g_L^g)
+6.2\mathrm{Re}(g_R^{Z*}g_R^g)]\left(\frac{\kappa^Z_{\mathrm
{tq}}}{\Lambda}\right)
\left(\frac{\kappa^g_{\mathrm{tq}}}{\Lambda}\right)\times10^
{-2}>1.4\times10^{-3} \mathrm{TeV}^{-2}\notag\\
\end{eqnarray}
for the integrated luminosity of $100\mathrm{fb}^{-1}$,
respectively.

\section{Conclusions}

In order to perform consistent studies for both the top quark
production and decay via FCNC couplings, we have calculated the NLO
QCD corrections to the three decay modes of the top quark induced by
model-independent FCNC couplings of dimension-five operators. For
$t\rightarrow q+g$, the NLO results increase the experimental
sensitivity to the anomalous couplings. Our results show that the
NLO QCD corrections enhance the LO branching ratio by about $20\%$,
as presented in Ref.~\cite{Zhang:2008yn}. Moreover, the NLO QCD
corrections vastly reduce the dependence on the renormalization
scale, which leads to increased confidence in our theoretical
predictions based on these results. For $t\rightarrow q+\gamma$ and
$t\rightarrow q+Z$, the NLO corrections are minuscule in branching
ratios, albeit they can decrease the LO widths by about $9\%$.
However, if we further consider the effects induced from operator
mixing, they can either be large or small, and increase or decrease
the branching ratios for $t\rightarrow q+\gamma$ and $t\rightarrow
q+Z$, depending on the values of the anomalous couplings
($\kappa^{g, \gamma, Z}_{\mathrm{tq}}/\Lambda$,
$f^{g,\gamma,Z}_{\mathrm{tq}}$ and $h^{g,\gamma,Z}_{\mathrm{tq}}$).

\begin{acknowledgments}
This work was supported in part by the National Natural Science
Foundation of China, under Grants No.10721063, No.10575001 and
No.10635030. C.P.Y acknowledges the support of the U.S. National
Science Foundation under Grand No. PHY-0855561. TCY was supported in
part by the National Science Council of Taiwan under Grant No.
98-2112-M-001-014-MY3.
\end{acknowledgments}

\textit{Note added.} After we completed this work we became aware of
a paper by Jure Drobnak, Svjetlana Fajfer and Jernej F. Kamenik
(arXiv:1004.0620) \cite{Drobnak:2010wh}, where they consider the
same effects from FCNC operator mixing.

\begin{figure}
\scalebox{0.7}{
\includegraphics{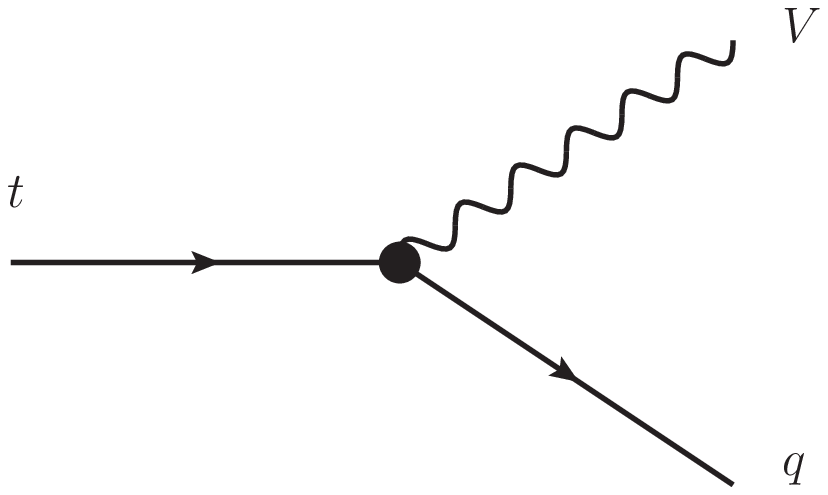}}
\caption{\label{tree} Tree-level Feynman diagram for $t\rightarrow
q+V$. Here, we use $q$ to represents the up-quark and the
charm-quark.}
\end{figure}

\begin{thebibliography}{}
\bibitem{Abazov:2007ev}
  V.~M.~Abazov {\it et al.}  [D0 Collaboration],
  Phys.\ Rev.\ Lett.\  {\bf 99}, 191802 (2007)
  [arXiv:hep-ex/0702005].

\bibitem{Aaltonen:2008qr}
  T.~Aaltonen {\it et al.}  [CDF Collaboration],
  Phys.\ Rev.\ Lett.\  {\bf 102}, 151801 (2009)
  [arXiv:0812.3400 [hep-ex]].

\bibitem{Liu:2005dp}
  J.~J.~Liu, C.~S.~Li, L.~L.~Yang and L.~G.~Jin,
  Phys.\ Rev.\  D {\bf 72}, 074018 (2005)
  [arXiv:hep-ph/0508016].

\bibitem{Yang:2006gs}
  L.~L.~Yang, C.~S.~Li, Y.~Gao and J.~J.~Liu,
  Phys.\ Rev.\  D {\bf 73}, 074017 (2006)
  [arXiv:hep-ph/0601180].

\bibitem{Carvalho:2007yi}
  J.~Carvalho {\it et al.}  [ATLAS Collaboration],
  Eur.\ Phys.\ J.\  C {\bf 52}, 999 (2007)
  [arXiv:0712.1127 [hep-ex]].

\bibitem{Han:1996ep}
  T.~Han, K.~Whisnant, B.~L.~Young and X.~Zhang,
  Phys.\ Rev.\  D {\bf 55}, 7241 (1997)
  [arXiv:hep-ph/9603247].

\bibitem{Han:1996ce}
  T.~Han, K.~Whisnant, B.~L.~Young and X.~Zhang,
  Phys.\ Lett.\  B {\bf 385}, 311 (1996)
  [arXiv:hep-ph/9606231].

\bibitem{Beneke:2000hk}
  M.~Beneke {\it et al.},
  arXiv:hep-ph/0003033.

\bibitem{Han:1995pk}
  T.~Han, R.~D.~Peccei and X.~Zhang,
  Nucl.\ Phys.\  B {\bf 454}, 527 (1995)
  [arXiv:hep-ph/9506461].

\bibitem{Hosch:1997gz}
  M.~Hosch, K.~Whisnant and B.~L.~Young,
  Phys.\ Rev.\  D {\bf 56}, 5725 (1997)
  [arXiv:hep-ph/9703450].

\bibitem{Obraztsov:1997if}
  V.~F.~Obraztsov, S.~R.~Slabospitsky and O.~P.~Yushchenko,
  Phys.\ Lett.\  B {\bf 426}, 393 (1998)
  [arXiv:hep-ph/9712394].

\bibitem{Abe:1997fz}
  F.~Abe {\it et al.}  [CDF Collaboration],
  Phys.\ Rev.\ Lett.\  {\bf 80}, 2525 (1998).

\bibitem{Han:1998tp}
  T.~Han, M.~Hosch, K.~Whisnant, B.~L.~Young and X.~Zhang,
  Phys.\ Rev.\  D {\bf 58}, 073008 (1998)
  [arXiv:hep-ph/9806486].

\bibitem{del Aguila:1999ec}
  F.~del Aguila and J.~A.~Aguilar-Saavedra,
  Nucl.\ Phys.\  B {\bf 576}, 56 (2000)
  [arXiv:hep-ph/9909222].

\bibitem{Chikovani:2000wi}
  L.~Chikovani and T.~Djobava,
  arXiv:hep-ex/0008010.

\bibitem{Kidonakis:2003sc}
  N.~Kidonakis and A.~Belyaev,
  JHEP {\bf 0312}, 004 (2003)
  [arXiv:hep-ph/0310299].

\bibitem{Chekanov:2003yt}
  S.~Chekanov {\it et al.}  [ZEUS Collaboration],
  Phys.\ Lett.\  B {\bf 559}, 153 (2003)
  [arXiv:hep-ex/0302010].

\bibitem{AguilarSaavedra:2004wm}
  J.~A.~Aguilar-Saavedra,
  Acta Phys.\ Polon.\  B {\bf 35}, 2695 (2004)
  [arXiv:hep-ph/0409342].

\bibitem{Zhang:2008yn}
  J.~J.~Zhang, C.~S.~Li, J.~Gao, H.~Zhang, Z.~Li, C.~P.~Yuan and T.~C.~Yuan,
  Phys.\ Rev.\ Lett.\  {\bf 102}, 072001 (2009)
  [arXiv:0810.3889 [hep-ph]].

\bibitem{Li:1990qf}
  C.~S.~Li, R.~J.~Oakes and T.~C.~Yuan,
  Phys.\ Rev.\  D {\bf 43}, 3759 (1991).

\bibitem{Greub:1997hf}
  C.~Greub and T.~Hurth,
  Phys.\ Rev.\  D {\bf 56}, 2934 (1997)
  [arXiv:hep-ph/9703349].

\bibitem{Amsler:2008zz}
  C.~Amsler {\it et al.}  [Particle Data Group],
  Phys.\ Lett.\  B {\bf 667}, 1 (2008).

\bibitem{Drobnak:2010wh}
  J.~Drobnak, S.~Fajfer and J.~F.~Kamenik,
  arXiv:1004.0620 [hep-ph].
\end{thebibliography}
\end{document}